\documentclass[usenatbib]{mn2e}
\usepackage[utf8]{inputenc}
\usepackage{natbib,epsfig,cite,mystyle,amsmath,amssymb,graphicx}
\usepackage[font=scriptsize]{caption}

\title[Evolution of the gas mass fraction in galaxy clusters]{Evolution of the
gas mass fraction in galaxy clusters}
\author[I. Dvorkin, Y. Rephaeli]{Irina Dvorkin$^{1,2}$\thanks{E-mail:
irina@wise.tau.ac.il}, Yoel Rephaeli$^{2,3}$\\
$^{1}$Institut d’Astrophysique de Paris, UMR 7095, 98 bis boulevard Arago,
Paris, 75014, France\\
$^{2}$School of Physics and Astronomy, Tel Aviv University, Tel Aviv, 69978,
Israel\\
$^{3}$Center for Astrophysics and Space Sciences, University of California,
San Diego, La Jolla, CA 92093-0424\\}

\begin{document}

\pagerange{\pageref{firstpage}--\pageref{lastpage}} \pubyear{2015}
\maketitle
\label{firstpage}

\begin{abstract}
 
The mass fraction of hot gas in clusters is a basic quantity whose level and 
dependence on the cluster mass and redshift are intimately linked to all cluster
X-ray and SZ measures. Modeling the evolution of the gas fraction is clearly a
necessary ingredient in the description of the hierarchical growth of clusters
through mergers of subclumps and mass accretion on the one hand, and the
dispersal of gas from the cluster galaxies by tidal interactions, galactic
winds, and ram pressure stripping on the other hand. A reasonably complete
description of this evolution can only be given by very detailed hydrodynamical 
simulations, which are, however, resource-intensive, and difficult to implement
in the mapping of parameter space. A much more practical approach is the use of
semi-analytic modeling that can be easily implemented to explore a wide range of
parameters. We present first results from a simple model that describes 
the build up of the gas mass fraction in clusters by following the overall
impact of the above processes during the merger and accretion history of each
cluster in the ensemble. Acceptable ranges for model parameters are deduced
through comparison with results of X-ray observations. Basic implications of our
work for modeling cluster statistical properties, and the use of these
properties in joint cosmological data analyses, are discussed.

\end{abstract}

\begin{keywords}
galaxies: clusters: general - intracluster medium
\end{keywords}

\section{Introduction}

Hot intracluster (IC) gas is an important cluster component that determines
X-ray emission quantities and the nature and properties of the
Sunyayev-Zel'dovich (SZ) effect.
Cluster X-ray and SZ surveys 
provide a broad basis for exploring key statistical properties of the
population, such as 
the mass function, and are valuable cosmological probes of, e.g., 
the equation of state of dark energy, the amplitude of primordial density
fluctuations and the neutrino mass \citep[e.g.][]{2009ApJ...692.1060V,
2010MNRAS.406.1759M, 2010ApJ...719.1045L, 2012MNRAS.427..828S,
2014A&A...571A..20P}. However, both 
X-ray emission and the SZ signal of a cluster of a given mass are very sensitive
to the hot gas mass fraction $f_{gas}$, which is not known precisely 
and - in principle - 
can depend on the mass and redshift of the cluster. While it is expected that
$f_{gas}$ should 
be close to the cosmic value $\Omega_b/\Omega_m$ by virtue of the large size of
clusters, 
some of the baryons are locked in cluster galaxies, and therefore do not
contribute to the 
respective observable. 
Therefore, it is of interest 
to model the fraction of hot, X-ray emitting gas in galaxy clusters,
particularly at high 
redshifts.

Observational effort to determine $f_{gas}$ is motivated also by the basic need 
to study the evolution of 
the total baryon fraction in groups and clusters, which has contributions also
from galaxies and 
IC light \citep[e.g.][]{1993Natur.366..429W,1999ApJ...517..627M,
2003MNRAS.344L..13E,2003ApJ...591..749L,2007ApJ...666..147G,2009ApJ...703..982G,
2007MNRAS.377.1457M}. In several of these works the reported baryon
fraction is smaller than the expected value, particularly in low-mass systems.
The observed trend is an increase of the fraction of hot gas with total system
mass, approximately 
following $f_{gas}\propto M^{0.1-0.2}$, and a decrease of the stellar fraction
as $f_s\propto M^{-(0.4-0.6)}$ \citep{2003ApJ...591..749L,2007ApJ...666..147G,
2009ApJ...693.1142S,2010ApJ...719..119D}.  
A possible interpretation of the mass dependence of $f_{gas}$ is that gas is
expelled from
low-mass systems due to non-gravitational processes, such as feedback from
active galactic nuclei (AGN) \citep{2004ApJ...608...62S}. In this scenario more
massive systems 
retain a larger fraction of their gas due to their deeper potential wells. 

Another important piece of evidence is the observed metallicity of IC 
gas, with a mean value of $\simeq 1/3$ solar 
\citep[e.g.][]{2000ApJ...544..188F,
2001ApJ...551..153D, 2005ApJ...628..655V, 2012A&A...537A.142B}, 
and with a decreasing radial profile. 
Since metals are produced only in stars, it follows that a large fraction of 
IC gas was ejected from galaxies. Indeed, numerical simulations
\citep[e.g.][]{2007A&A...466..813K,2010ApJ...716..918A} show that 
ejection of metals from galaxies can account for the observed metallicity. 
This interpretation is further strengthened by the observed evolution of the 
galaxies in 
clusters (the Butcher-Oemler 
effect), namely that 
the fraction of blue galaxies is higher at higher redshifts 
\citep{1978ApJ...219...18B}; note though that 
the significance of this effect is 
uncertain due to difficulties in disentangling the influence of the chosen
galaxy sample and secular evolution \citep{2012A&A...543A..19R}. 
Moreover, spirals found in clusters tend to be redder
\citep{2009MNRAS.396L..41H}, H\textsc{I} deficient as compared to 
similar galaxies in the field \citep{2001ApJ...548...97S}, and 
typically have truncated 
gaseous discs \citep{2006AJ....131..716K}.
These observations suggest that galaxies lost most or all of their gas
since they first fell into the cluster, either due to encounters with other
galaxies, or as a result of ram-pressure stripping, 
as discussed below. This led to the quenching of star formation and the
subsequent change of color and morphology. 

The baryonic fraction in clusters, particularly the fraction of hot gas, 
was extensively studied using cosmological hydrodynamical simulations
\citep[e.g.][]{2006MNRAS.365.1021E, 2006MNRAS.367.1641B, 2010ApJ...715.1508S,
2011MNRAS.413..691Y,
2013MNRAS.tmp..918P}. While the observed mass dependence of $f_{gas}$ is
generally
well reproduced by these simulations, the stellar mass fraction is typically
larger than observed, which may be due to the fact that the (correct) gas mass
fraction is attained by an overestimated star formation rate. In general,
numerical simulations are computationally expensive, which complicates the 
modeling of the interplay between galactic and large-scale phenomena.

An alternative semi-analytic model, proposed by \citet{2009ApJ...700..989B}
\citep[see also][]
{2005ApJ...634..964O}
is based on the assumption that $f_{gas}$ in all halos was initially equal to
the cosmic 
baryon fraction, and that it 
decreased due to the processes of star formation 
and ejection of gas out of the halo by SN-and-AGN-driven winds. 
This model is calibrated to X-ray observations of nearby clusters, and so
successfully reproduces the local cluster population. 

In this paper we take a different 
approach: motivated by the observed metallicity and galaxy (color) evolution, we
assume 
that a large fraction of the IC gas was ejected from galaxies. In this picture
$f_{gas}$ 
increases with mass because larger systems typically form later through mergers
of smaller
systems, and therefore their galaxies had more time to eject their gas. As we
show in this paper, 
our model directly links galactic processes (which can be described by 
small-scale numerical simulations) with various cluster-scale observables.

Several processes may be responsible for mass ejection from cluster galaxies.
When a galaxy traverses the higher density inner region of a cluster,
ram-pressure effectively removes an appreciable fraction of its interstellar
gas \citep{1972ApJ...176....1G}. The details of this process depend on the
IC gas density profile, galactic gas density profile, and the 
trajectory of the galaxy 
\citep{1999MNRAS.308..947A, 2001ApJ...561..708V,
2006ApJ...647..910H, 2010MNRAS.408.2008T}, all of
which are difficult to model, but it is clear that ram pressure can remove
large 
quantities of gas from the galaxy on relatively short timescales 
\citep[e.g.][]{2000Sci...288.1617Q}. 
Observational evidence for this process comes from the tails behind several
cluster galaxies seen in H\textsc{I}, H$\alpha$ and X-rays, interpreted as
removal of galactic interstellar medium (ISM)
\citep[e.g.][]{2007ApJ...671..190S,2014ApJ...781L..40E}.
In addition, tidal interactions 
between field galaxies are known to affect the distribution of gas and stars
within the galaxies, and may be
as important in cluster galaxies, especially in 
dense cluster cores \citep{1983ApJ...264...24M, 1999MNRAS.304..465M,
2003ApJ...582..141G}. 
Tidal interactions truncate the dark matter density profile of 
subhalos orbiting inside a massive cluster, which leads to more concentrated
profiles of subhalos relative to field halos \citep{2001MNRAS.321..559B,
2009ApJ...696.1771L}. 
The transformation of spirals into S0 galaxies in clusters and the existence of
'passive spirals' (which are morphologically identical to normal spirals but
lack star formation activity) may be related to these environmental effects
\citep[e.g.][]{2002ApJ...577..651B, 2010ApJ...711..192J}. 

Other major processes that affect 
galaxy evolution in clusters are galactic winds and AGN feedback. 
SN-driven winds are particularly important at high redshifts, when the star
formation rate (SFR) is high \citep[e.g.][]{2001ApJ...554..981P}. 
A sufficiently fast wind deposits metal-enriched material into 
IC space. However, as the SFR drops at low redshifts, metal enrichment 
effects of galactic winds become sub-dominant to ram pressure and tidal
interactions. 

Outflows launched from the vicinity of supermassive black holes found in the
central galaxies of clusters may provide the IC gas with enough energy to
escape the cluster potential well and thus lower the IC gas mass fraction
\citep[see e.g.][for a comprehensive discussion of the role of AGN in
clusters]{2012ARA&A..50..455F}. X-ray cavities in the central regions of many
clusters, often found in pairs, provide strong evidence of these outflows. Their
ubiquity is, however, a matter of debate. The detection
fraction of X-ray cavities is between $20\%$ and $30\%$ in X-ray selected
samples \citep{2004ApJ...607..800B,2005MNRAS.364.1343D,2012MNRAS.421.1360H}. On
the other hand, a recent study by \citet{2014arXiv1410.0025H} found only $6$
clusters with X-ray cavities in a sample of $83$ massive clusters selected by
their SZ signature. While the detection of X-ray cavities is observationally
challenging due to their small contrast with the surrounding medium, their
contribution to the overall properties of the cluster population (rather than
individual clusters with extremely strong outflows) remains an open question.
Moreover, while the energy deposited into the IC gas is probably sufficient to
prevent overcooling \citep{2007ARA&A..45..117M} it is not clear that large
amounts of gas can also escape the potential well of the clusters.

In view of these findings we focus here on environmental processes which are
widespread in galaxy clusters and are closely related to their mass accretion
histories. 

In this paper we build a phenomenological model of gas ejection in the context
of the hierarchical assembly of clusters and explore the range of possible
models and their consequences for X-ray and SZ cluster surveys.
We adopt the following cosmological
parameters: $H_0=67.11$ km/s/Mpc, $\Omega_m=0.3175,\: \Omega_{\Lambda}=0.6825$,
$\sigma_8=0.8344$, and $n_s=0.9624$ \citep{2014A&A...571A..16P}. 
Unless
overwise stated, all masses $M$ and radii $R$ represent the virial quantities,
defined by $M=4\pi/3\: \Delta_V\rho_c(z) R^3$ where $\rho_c$ is the critical
density of the Universe at redshift $z$ and $\Delta_V$ is the overdensity
defined by the spherical collapse model \citep{1972ApJ...176....1G} calculated
for the given set of cosmological parameters.

In Section 2 we briefly describe our model of cluster evolution, which is based
on an extended merger tree code that follows the evolution of halos that consist
of dark matter and baryons. Gas ejection from galaxies and the build-up 
of $f_{gas}$ is discussed in Section 3. Our results are presented in Section 4
and discussed in Section 5.

\section{Cluster merger-tree evolution}
\label{sec:mtree}

The efficiency of interstellar (IS) gas removal by tidal interactions and ram
pressure depends on the depth of the cluster gravitational potential. This
occurs continuously through a series of interaction and merger events during
the dynamical evolution of a galaxy in a growing cluster. We follow the
evolutionary history of IC gas by considering the overall impact of the above
galactic processes in a statistical description based on a merger tree code.

In the $\Lambda$CDM framework structure forms hierarchically, starting with
relatively low mass halos that grow successively through mergers and accretion.
The merger history of a 
given cluster can be described by a \emph{merger tree}, which 
essentially is a list of the masses of the merging halos and the redshifts at
which these mergers occurred. 
The mass assembly history of a cluster affects its density profile
\citep{2002ApJ...568...52W,2010arXiv1010.2539D} and causes an intrinsic scatter
in all the
mass-observable relations. In previous works \citep{2011MNRAS.412..665D,
2012MNRAS.421.2648D} we studied how the 
hierarchical formation of galaxy clusters affects their X-ray and SZ 
properties; here we 
extend our merger-tree approach to include 
an approximate description of some basic galaxy-scale processes.

In order to describe the evolution of galaxy clusters we build
merger trees of dark matter halos using the \textsc{galform} algorithm
\citep{2008MNRAS.383..557P} which is based on the excursion set formalism
\citep{1993MNRAS.262..627L}. For a cluster with a given mass and at a given
observation redshift 
each merger tree represents a possible formation history, and a sufficiently
large number of merger trees 
can provide a statistical description of the population. 
The advantage of using this kind of semi-analytic modeling is our ability to
produce a large number of clusters (equivalent to simulating a very large
volume of the Universe) by employing an efficient algorithm that can be readily
applied to explore the parameter space of the model.

Instead of using a constant redshift step for the output of the merger tree, we
save the information on all the progenitors with masses 
$M>M_{res}$, where $M_{res}$ is the mass resolution limit. 
The merger trees of \citet{2008MNRAS.383..557P} are calibrated to match the 
Sheth-Tormen mass function \citep{1999MNRAS.308..119S}, which we use throughout
this paper for consistency. 
The original DM-only merger-tree code was extended to include also IC gas, 
whose density and temperature profiles are determined from basic considerations 
(essentially, energy conservation and hydrostatic equilibrium). 
Further details on the merger tree algorithm and its implementation for clusters
of galaxies can be found in \citet{2008MNRAS.383..557P} and 
\citet{2011MNRAS.412..665D}.

We follow the evolution of all 
halos in a tree (i.e. all the progenitors of the given cluster) with (total)
masses $M>M_{res}=10^{11}M_{\odot}h^{-1}$ that existed below redshift $z=2$. 
The number of galaxies in a halo scales linearly with its mass $M$, 
therefore
we calculate the initial number of galaxies in each halo as
\begin{equation}
 N_{gal,i}(M)=N_{gal,0}\left(\frac{M}{10^{11}M_{\odot}h^{-1}}\right)
\label{eq:Ngal}
\end{equation}
where $N_{gal,0}$ is a model parameter.
At high enough redshift large
structures
are 
rare; therefore, their member galaxies 
are expected to resemble low-redshift galaxies in the field, i.e. they should be
relatively 
massive blue disks. We assign an initial mass for these galaxies
$M_{gal,i}=10^{11}M_{\odot}h^{-1}$, a value that decreases by the various mass
loss processes. 

An alternative method of describing the galaxy population would be to explicitly
account for 
subhalos and follow them as distinct 
systems even after they merge with the main halo. This kind of approach
\citep[i.e.][]{2007ApJ...667..813Y} necessitates modeling the trajectory of
each galaxy inside the main halo, taking into account dynamical 
friction, encounters with other subhalos, and the impact of subsequent mergers
of the main halo with other systems. While this kind of approach provides a more
accurate description of cluster growth, it might be difficult to pinpoint the
key physical processes that influence the evolution of 
IC gas. Therefore, we chose to assign all the galaxies 
the same (fiducial) initial mass, which is reduced at later stages of evolution
as described below. We note that our model effectively averages over all
possible galaxy masses and trajectories, 
as well as the merger impact parameters.

The gas mass fraction in the diffuse matter that was not contained in collapsed
structures (this gas could originate from early
galactic winds) is $f_{diff}$ at the initial time, so that
$M_{gas,i}=f_{diff}\cdot (M-M_{gal,i}N_{gal})$, whereas galaxies 
are assumed to have 
the cosmic baryon fraction $f_c=\Omega_b/\Omega_m$. 
Clearly, baryon ejection processes from galaxies affect the stellar component, 
the disk, and the warm gas in the galactic halo.

\section{Modeling IC environmental processes}

When galaxies fall onto larger structures, they experience tidal interactions 
with the host halo and other subhalos. The strength of these interactions
depends on the host halo mass, or on the local density of galaxies, which is
ultimately also determined by the host mass. Tidal interactions affect the dark
matter, as well as IS gas, so that cluster galaxies are expected to have
truncated dark matter profiles \citep{2009ApJ...696.1771L}. This truncation
enhances the effect of ram pressure stripping by making the galaxy potential
wells effectively shallower.

Numerical simulations show that the time scale of gas removal from galaxies
through 
ram pressure is relatively short \citep{2000Sci...288.1617Q}, and that the
fraction of 
gas-depleted galaxies is a strong function of host halo mass
\citep{2006ApJ...647..910H, 
2010MNRAS.408.2008T}. Thus, although the ram pressure experienced by a galaxy 
moving in a cluster varies between a maximal value attained at the cluster
center and a minimal value in the outskirts \citep{2008MNRAS.383.1336B}, 
most loosely bound gas is likely to be ejected upon first passage through the
center. Further stripping occurs when the galaxy 
is in a deeper potential well, i.e. after a merger with a larger halo. This
episodic mass loss and the connection between ram pressure stripping and the
merging history of the cluster is demonstrated by the numerical simulations of
\citet{2007A&A...466..813K}. On the other hand, the model by
\citet{2006ApJ...647..910H} shows that for a galaxy of 
a given mass there exists a limiting cluster mass for which 
ram pressure is strong enough to remove almost all of the gas even in the
innermost regions of the galaxy, while for smaller cluster masses almost none of
the gas is removed. 

Motivated by these findings, we assume that the efficiency of gas removal is a
power-law in cluster mass. If this dependence is steep, all available gas will
be removed from the galaxies once the cluster reaches a certain mass, 
resulting in gas mass fraction which is roughly a step-function in cluster
mass. 
On the other hand, if the efficiency is only weakly dependent on cluster mass, 
gas will be removed from galaxies in a more gradual manner. We stress that this 
is a phenomenological model intended to link the observable properties of galaxy
clusters to galactic processes. In other words, our model parameters, in particular 
those related to the efficiency of gas removal, 
provide the basis for an effective 
description of more complex galactic processes. We briefly discuss how
our model can be used to study these processes in Section \ref{sec:discussion}.

Our calculation proceeds as follows. After each merger event a fraction 
\begin{equation}
 f_m=f_{m,0}\left(\frac{M}{10^{14}M_{\odot}h^{-1}}\right)^{\alpha}
\label{eq:fm}
\end{equation}
of the total galactic mass is removed, of which $f_c$ is in baryons:
\begin{equation}
 \Delta M_{gas}=f_m f_c N_{gal}M_{gal}\:.
\label{eq:deltamgas}
\end{equation}
This gas is no longer bound to the galaxy and is assumed to immediately mix
with the IC gas.
In eq. (\ref{eq:deltamgas}), $M$ is the halo mass, $N_{gal}$ is given in
equation (\ref{eq:Ngal}) and
$M_{gal}$ is the mass of a typical galaxy residing in 
this halo. 
This mass is reduced after each merger event to account for the mass 
loss as follows:
\begin{equation}
 M_{gal}\rightarrow M_{gal}\cdot(1-f_m)\:
\end{equation}
The parameter $\alpha$ describes the steepness of the dependence of gas removal
on cluster mass; for large values of $\alpha$ there will be a very pronounced
transition from insignificant 
environmental effects to very rapid 
mass ejection 
from the galaxy, whereas for small values of $\alpha$ 
the ejection process is more gradual.

We approximate the virialization phase of gas removed from galaxies by assuming 
that it is is immediately heated 
to the virial temperature of the host halo. 
Thus, the hot gas content of a halo immediately after a merger event is 
\begin{equation}
 M_{gas}(M)=M_{gas}(M_1)+M_{gas}(M_2)+\Delta M_{gas}+\Delta M_{gas,diff}
\end{equation}
where $M_1,M_2$ are the masses of the two merging halos (typically the host halo
and a smaller halo), $\Delta M_{gas}$ is calculated as in eq.
(\ref{eq:deltamgas}),
and $\Delta M_{gas,diff}$ is the gas contained in diffuse matter that falls onto
the halo. 
This last term is given by $\Delta M_{gas,diff}=f_{diff}M_{diff}$ where
$M_{diff}$ is the mass contained in halos below the resolution limit $M_{res}$
of the merger tree (see Section \ref{sec:mtree}).

These equations are employed following each merger event, so that the gas
content of the cluster increases as the cluster 
evolves and the mean mass of cluster galaxies is decreased.

\section{Results}

\subsection{IC gas fraction and metallicity}

We ran $1000$ tree realizations for each halo
mass with mass resolution of $M_{res}=10^{11}M_{\odot}h^{-1}$ and up to $z=2$.
We considered all merger events between halos above this resolution mass.
Mergers with smaller halos were considered as part of the smooth accretion
process, as described above.

\begin{figure}
\centering
\epsfig{file=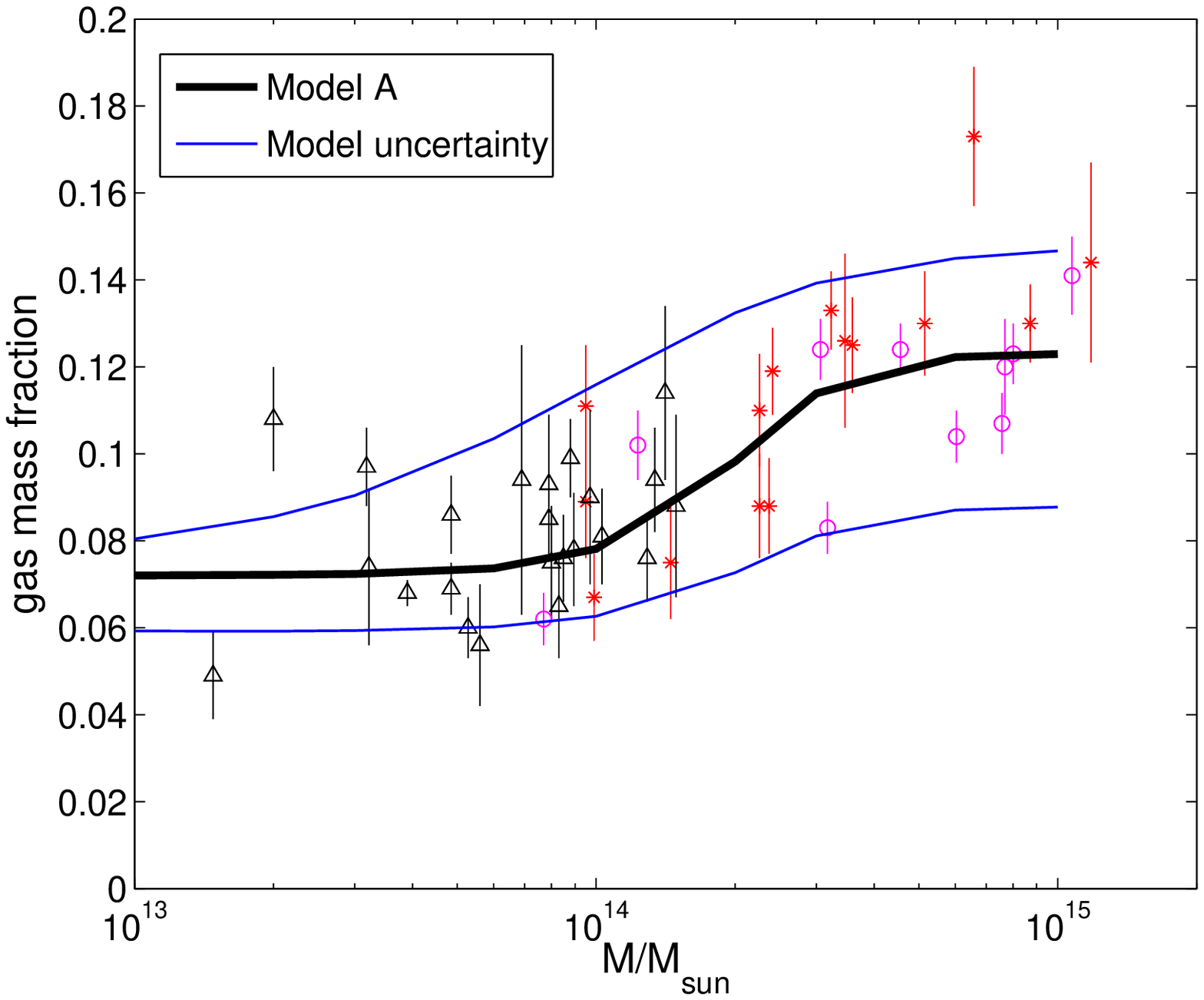, height=6cm}
\caption{Gas mass fraction vs. cluster 
mass for the fiducial model (thick black line) with model uncertainty range
bracketed
by the thin blue lines. Paramater values are specified in Table
\ref{tab:modparams}. X-ray measurements inside
$M_{500}$ from \citet[][red stars]{2013ApJ...778...14G}, \citet[][magenta
circles]{2006ApJ...640..691V} and \citet[][black
triangles]{2009ApJ...693.1142S}.}
\label{fig:fgas_main}
\end{figure}

Fig. \ref{fig:fgas_main} shows the hot gas mass fraction in clusters in the mass
range $10^{13}M_{\odot}h^{-1}-10^{15}M_{\odot}h^{-1}$ as predicted by our model,
and
compared with data from several X-ray studies. 
Parameters of the fiducial model, shown in black, are
given in Table \ref{tab:modparams}.
We compare our model
with X-ray observations of groups and clusters 
(points with error-bars) and find good agreement. These results suggest that the
mass dependence of $f_{gas}$ can be 
largely explained by environmental processes. 
To obtain an estimate for the range of parameter values that are consistent with
the data, we show by the thin blue lines in Fig. \ref{fig:fgas_main} the
approximate region that brackets the range of values of the three datasets.
These upper and lower lines correspond to
models $A_{max}$ and $A_{min}$
respectively, whose parameters are given in Table \ref{tab:modparams}. 
Note though that the comparison with observations has only a
limited value due to substantial modeling uncertainty, mainly in the cluster
mass determination from X-ray observables.

\begin{table}
\centering
\begin{tabular}[h]{| c || c | c | c | c | c |}
 \hline
               & $A$    & $B$    & $C$    & $A_{min}$ & $A_{max}$ \\ \hline
  \hline
  $N_{gal,0}$  & $0.5$  & $0.3$  & $0.5$  & $0.3$     & $0.7$     \\ \hline
  $f_{diff}$   & $0.11$ & $0.12$ & $0.12$ & $0.0725$  & $0.14$    \\ \hline
  $f_{m,0}$    & $0.01$ & $0.01$ & $0.01$ & $0.01$    & $0.045$   \\ \hline
  $\alpha$     & $1.84$ & $1.0$  & $0.7$  & $1.55$    & $0.39$    \\
 \hline

\end{tabular}
\caption{Parameter values used for the fiducial model $A$ (solid black curve
in Figs. \ref{fig:fgas_main}, \ref{fig:fgas_models}, \ref{fig:fgas_metals});
representative models $B$ and $C$ (dashed red curve and dot-dashed magenta
curve, respectively), and the lower and upper ranges of the models, $A_{min}$
and $A_{max}$, depicted as thin blue lines in Fig. \ref{fig:fgas_main}.}
\label{tab:modparams}
\end{table}

An analytical fit to the fiducial model is
\begin{equation}
 f_{gas}=c\left(1+e^{-\left[\log_{10}(M/M_{\odot})-a\right]/b} \right)^{-1}+d
\:.
\label{eq:modelA}
\end{equation}
The fit parameters for the models shown on Figure \ref{fig:fgas_main} are given
in Table \ref{tab:modfits}.
The fit demonstrates the transition to efficient 
mass stripping, which occurs around $M=10^a M_{\odot}$.

No explicit redshift dependence is deduced 
from our results, in line with
the usual assumptions \citep[e.g.][]{2008MNRAS.383..879A}.
The reason for this is that in our model the efficiency of gas removal from
cluster galaxies depends on the mass of the host 
halo, but not explicitly on the redshift at which galaxy infall 
occurs. Note though that the dependence on mass clearly introduces implicit
redshift dependence through the strong mass dependence of the probability
distribution function of cluster formation 
times \citep[e.g.][]{2008MNRAS.388.1759S}.

\begin{figure}
\centering
\epsfig{file=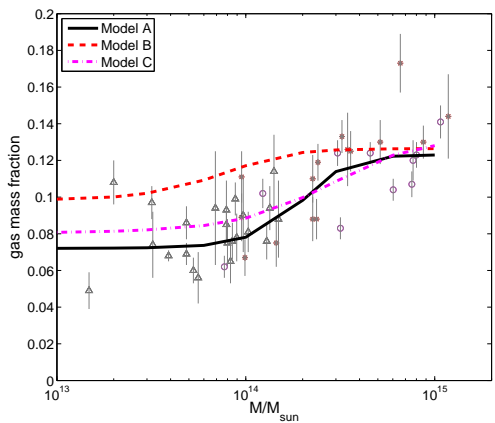, height=6cm}
\caption{Gas mass fraction vs. cluster mass for $3$ representative models (see
Table \ref{tab:modparams} for the parameters used in each case). 
Points with error bars show the X-ray measurements, 
specified in the caption of Figure \ref{fig:fgas_main}.}
\label{fig:fgas_models}
\end{figure}

 \begin{table}
\centering
  \begin{tabular}[h]{| c || c | c | c | c | c |}
 \hline
               & $A$    & $B$    & $C$    & $A_{min}$ & $A_{max}$ \\ \hline
  \hline
  a            & 14.29  & 13.87  & 14.45  & 14.32     & 13.95     \\ \hline
  b            & 0.135  & 0.189  & 0.251  & 0.148     & 0.295     \\ \hline
  c            & 0.051  & 0.028  & 0.053  & 0.029     & 0.071     \\ \hline
  d            & 0.072  & 0.099  & 0.081  & 0.059     & 0.078     \\ \hline
  \end{tabular}
  \caption{Parameter values for the fit in eq. (\ref{eq:modelA}) for the
different models discussed in the text. Note that the physical parameters that
define the models are given in Table \ref{tab:modparams}. The transition to
efficient mass stripping occurs at around $M=10^a M_{\odot}$ for each model.}
  \label{tab:modfits}
 \end{table}

Fig. \ref{fig:fgas_models} shows the results of our model for three
representative sets of parameters: Model A (the fiducial model shown on Fig.
\ref{fig:fgas_main}; solid black line), Model B (dashed red line) and Model C
(dot-dashed magenta line). The corresponding parameters are given in Table
$\ref{tab:modparams}$.
In all 
three models the gas mass fraction increases with mass by $\sim 20-50\%$ 
from groups to rich clusters, respectively. 
This trend is largely determined 
by the following model parameters: $N_{gal,0}$, which is related to the amount 
of gas initially locked inside galaxies, and $f_{diff}$, which is the gas
fraction 
in diffuse matter. The latter parameter is expected to be high, but lower than
the 
universal value $f_c$ since baryons are more clustered than 
dark matter. The parameters $f_{m,0}$ and $\alpha$ quantify the environmental 
processes that galaxies undergo as the cluster is assembled and affect the
transition 
from the low $f_{gas}$ level 
in groups to a high level in rich clusters. In particular, 
the steepness of this transition is determined by the value of $\alpha$ (see eq.
\ref{eq:fm}).

While a more complete and quantitative description of gas ejection processes
requires a high spatial resolution hydrodynamical simulation that can follow
individual galaxy trajectories, our simple treatment seems to provide an
adequate basis for comparison with the data. The good agreement of the
predicted mass dependence of $f_{gas}$ with 
the observations clearly indicates that \emph{on average} cluster environmental 
processes may be described by a few universal model parameters. 
Interestingly, the observed $f_{gas}$ exhibits large scatter, which may be
linked to 
the scatter in these unresolved parameters. Quantifying the connection between
the 
varying galactic trajectories, composition, IC gas density profile, 
values of the merger impact parameter, 
and the scatter in our effective model parameters, 
is an important future goal 
(which is clearly beyond the scope of our simplified treatment).

\begin{figure}
\centering
\epsfig{file=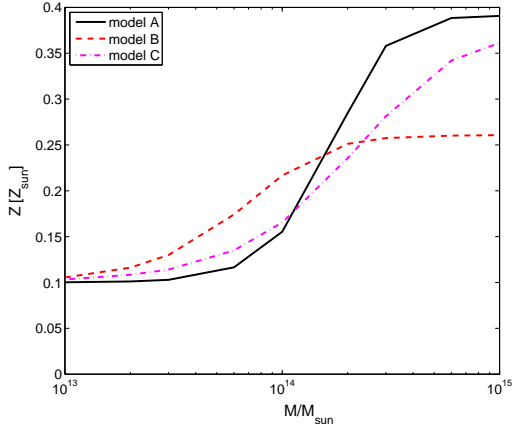, height=6cm}
\caption{Metal abundance (in units of the solar value) for 
models A, B and C
for which the corresponding $f_{gas}$ is shown 
in Fig. \ref{fig:fgas_models}.}
\label{fig:fgas_metals}
\end{figure}

IC gas metallicity provides additional insight on the evolution of the gas mass
fraction.
Gas that was removed from galaxies obviously has higher
metal abundance than the 
inter-cluster gas.
However, the metallicity of the latter which we denote by $Z_i$ 
could be higher than that of 
primordial gas, since it 
could have already been 
enriched by galactic outflows \citep{2013Natur.502..656W}. On the other hand,
the metallicity of galactic gas, $Z_{gal}$ depends on the stellar mass and
probably also 
on the environment of the galaxy \citep{2014MNRAS.438..262P}. Since these 
parameters depend on processes that occurred before $z=2$,
long before the cluster had assembled, we do not attempt to model them here,
instead we adopt effective constant values for both $Z_i$ and $Z_{gal}$.

Fig. \ref{fig:fgas_metals} shows the metallicity of the 
IC gas for various cluster masses with $Z_i=0.1$ and $Z_{gal}=0.8$. It can be
seen that although all three models produce similar $f_{gas}(M)$ they can be
further distinguished 
by their very different mass-metallicity relations (which, however,
depend on the assumed
ratio of $Z_{i}/Z_{gal}$). The differences between the selected models are
mostly evident for the most massive clusters, but also for $M\lesssim
10^{14}M_{\odot}$ which roughly corresponds to the halo mass for which
environmental effects become important. 
Interestingly,
there seems to be some observational evidence 
\citep{2009MNRAS.399..239R,2014ApJ...781...36S} for an increase of the mean 
metal abundance with IC gas temperature, 
and hence with system mass (although other studies, e.g.
\citet{2011A&A...535A..78Z,2012A&A...537A.142B} favor an
inverse trend). 
We caution, however, that current X-ray studies measure the metal abundance in
the innermost region of the cluster, typically inside $R_{500}$ or less, due to
observational difficulties, and the evolution of the
central metal abundance might differ appreciably from the evolution of the total
metallic mass in the cluster.
Further investigation of the chemical composition
of IC gas in groups and clusters will help constrain our model and 
provide more information 
on the processes of gas enrichment.

\subsection{X-ray luminosity}
\label{sec:xray}

Modeling $f_{gas}(M)$ is particularly important in view of the ongoing and
upcoming 
X-ray and SZ cluster surveys, whose main objectives are the study of cluster
properties and the use of clusters as precise cosmological probes. 
These surveys, when jointly analyzed together 
with complementary cosmological probes (such as 
CMB anisotropies and baryonic acoustic oscillations) can shed light on the
physics of galaxy clusters and the nature of mass-observable relations. The
latter are shaped by cosmological 
structure formation, as well as small-scale 
physics. 
Therefore, the magnitude of $f_{gas}$ and its $M \& \, z$
dependence may effectively serve as a means of studying 
the main physical processes affecting 
galaxies in dense environments, provided the cosmological parameters are
constrained
fairly well by complementary probes.
In particular, the $f_{gas}$ models proposed here can be used to link X-ray
observables to 
galactic processes in clusters.

First we calculate the scaling relation between the bolometric X-ray luminosity
and the emission weighted temperature inside $R_{500}$. 
In order to calculate the density and temperature
profile we assume a polytropic model 
\citep[e.g.][]{2003MNRAS.346..731A}, 
so that the pressure and density are
related by $P=P_0(\rho/\rho_0)^{\Gamma}$ with $\Gamma=1.2$. Then the solution of
the equation of hydrostatic equilibrium for a polytropic gas 
inside a potential well of a DM halo
with an NFW profile 
is \citep{2005ApJ...634..964O}:
\begin{equation}
\label{eq:rhogas}
 \rho(x)=\rho_0\left[1-\frac{B}{1+n}\left(1-\frac{\ln(1+x)}{x}
\right) \right]^{n} ,
\end{equation}
where $n=(\Gamma-1)^{-1}$, $B=4\pi G\rho_s r_s^2\mu m_p/k_B T_0$, $\mu m_p$ is
the mean molecular weight, $r_s=R/c$ is the scale radius of the 
cluster, and $c$ is the concentration parameter
of the DM halo. 
The temperature profile is given by:
\begin{equation}
\label{eq:tgas}
	T(x)=T_0\left[1-\frac{B}{1+n}\left(1-\frac{\ln(1+x)}{x} \right)
\right]\: .
\end{equation}
We take the halo concentration parameters from our merger-tree model, which
describes the dark matter density profile as a function of the formation
history of the cluster
\citep{2011MNRAS.412..665D, 2012MNRAS.421.2648D}. 

\begin{figure}
\centering
\epsfig{file=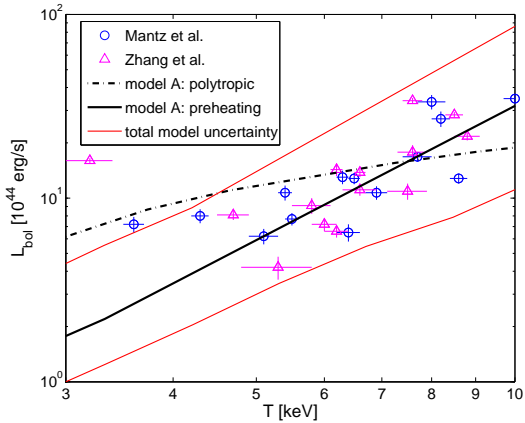, height=6cm}
\caption{
Bolometric luminosity - temperature relation computed using the
polytropic $f_{gas}$ model
(dot-dashed black line), preheating model with a uniform entropy floor of $150$
keV cm$^2$
(solid black line) and model uncertainty region (thin red lines) compared with
X-ray
measurements from \citet{2008A&A...482..451Z} and \citet{2010MNRAS.406.1773M}.}
\label{fig:xray_lowz}
\end{figure}

Figure \ref{fig:xray_lowz} shows the bolometric luminosity $L_X$ vs. emission
weighted temperature for model clusters at $z=0$ compared with low-redshift
observations with the polytropic model shown by the dot-dashed black line. While
the model
prediction reasonably agrees 
with the data, the slope of the luminosity-temperature relation is too shallow and 
follows the self-similar prediction $L\propto T^2$, instead of the observed 
$L\propto T^3$. This discrepancy clearly stems from our simplistic description of 
IC gas equation of state. 
Quite possibly this discrepancy indicates the need for additional 
energy input from supernovae, galactic winds or AGN
which are related to
the cluster 
galactic evolutionary processes. These lead also to the overall effect of gas 
preheating, as has been proposed in 
various preheating models 
\citep[e.g.][]{2001ApJ...555..597B,2002MNRAS.330..329B,2003ApJ...593..272V}.
This calculation is beyond the scope of the present paper and we leave it to
future work. However, in order to demonstrate the
effect of this preheating we include a simple model, based on the work of  
\citet{2007ApJ...666..647Y} and 
described in the Appendix, where we assume a constant entropy floor for all
clusters. This assumption amounts to uniformly raising the entropy level of all
intergalactic gas well before the formation of groups and clusters. 
Since possible sources of preheating are related to star formation, which peaks at
$z\sim 2$, this assumption is fairly reasonable. Nevertheless, this description
is not entirely self-consistent if energy injection 
continues 
also at lower redshifts, as we do not consider the effects of gas ejection from groups 
due to increased entropy, which 
could affect the evolution of 
the gas mass fraction. 
Indeed, there are observational hints for the connection between the gas mass fraction
and the entropy profile \citep{2010A&A...511A..85P}. 
(In Section \ref{sec:discussion} we briefly outline our plan for a more fully
consistent treatment of preheating in our merger-tree approach.)
 
The solid black curve in Figure \ref{fig:xray_lowz} shows the results of the
preheating model, where we assumed an entropy floor of $K_0=150$ keV cm$^2$.
This model clearly provides a much better fit to observations. Note, however,
that there is a significant dispersion in the observed entropy floor
\citep{2010A&A...511A..85P,2014ApJ...794...67M}. In the  
absence 
of a complete model that follows the development of this entropy excess in each 
individual cluster, we explore a plausible range of values for $K_0$.
The thin red curves in Figure \ref{fig:xray_lowz} 
mark the estimated range resulting from the uncertainty in the gas mass fraction
model (blue curves in Figure \ref{fig:fgas_main}) and in the entropy floor:
$K_0=100-200$ keV cm$^2$. This combined uncertainty brackets the observations,
as expected. The observed scatter in the X-ray scaling relations might be due to
different dynamical state of some of the clusters (i.e. they could be out of
equilibrium due to a recent merger event), as well as variations in $f_{gas}$. 

Having demonstrated the viability of the preheating assumption for our model
we use the simple polytropic case in the remainder of this paper in order to
isolate the effects of our IC gas model.

Additional information is provided by the cluster luminosity function.
Recently, \citet{2014A&A...570A..31B} used the REFLEX II cluster survey to
construct the X-ray luminosity function and to derive constraints on $\Omega_m$ and $\sigma_8$. Future surveys will be able to extend this analysis to higher redshifts, 
probing the mass function and thermodynamical properties of these systems. 

\begin{figure}
\centering
\epsfig{file=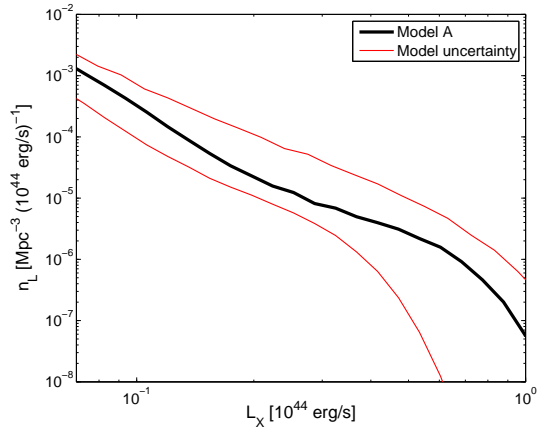, height=6cm}
\caption{X-ray luminosity function computed using the fiducial $f_{gas}$ model
(thick black line) and model uncertainty region (thin red lines) for clusters in
the range $z=0-0.5$. The luminosity is calculated for the $0.1-2.4$ keV spectral
band.}
\label{fig:lumifunc}
\end{figure}

Figure \ref{fig:lumifunc} shows the expected luminosity function for redshifts
up to $z=0.5$ in the 
measured $0.1-2.4$ keV spectral band, 
which corresponds to the 
energy range of ROSAT measurements. 
We used the Sheth-Tormen mass function for consistency with the merger tree
algorithm we employ. In the future we plan to extend
this work by calibrating the merger tree building code to more general mass
functions, so as to carry out a more detailed comparison with results of
hydrodynamical simulations.

\subsection{SZ power spectrum}

Recent findings by the \emph{Planck} Collaboration (2013) 
indicate that there is `tension' 
between the observed SZ power spectrum and cluster number counts and 
theoretical predictions based on primary CMB observations. One of the possible 
culprits is the gas mass fraction, which links the dark matter halo abundance, 
predicted by theory, to the observed SZ signal, which results from interaction
of CMB
photons and hot IC electrons. It is quite interesting, therefore, to check
whether our range of $f_{gas}$ models can alleviate the tension reported by
\emph{Planck}. 

We compute the SZ power spectrum using the halo approximation
\citep{2002MNRAS.336.1256K}:
\begin{equation}
 C_{\ell}=s(\chi)^2\int_0^{z_{max}}\frac{dV(z)}{dz}dz\int_{M_{min}}^{M_{max}}{
dM \frac { dn } { dM} |y_{\ell}(M,z)|^2}
\label{eq:cl}
\end{equation}
where $s(\chi)$ 
is the spectral dependence of the SZ signal given by:
\begin{equation}
  s(\chi)=\chi \frac{e^{\chi}+1}{e^{\chi}-1}-4,
\end{equation}
$\chi=h\nu/k_BT_0$ is the dimensionless frequency, 
$V(z)$ is the comoving volume per steradian, and $dn/dM$ is the 
mass function. 
The 2D Fourier transform of the projected Comptonization parameter is
\begin{equation}
 y_{\ell}=\frac{4\pi r_s}{\ell_s^2}\int_0^c{dx x^2 \frac{sin(\ell
x/\ell_s)}{\ell x/ \ell_s} \zeta(x)}
\end{equation}
where $\ell_s=d_A(z)/r_s$, $d_A(z)$ is the angular diameter distance to the
cluster, and $\zeta(x)$ is the gas (normalized) pressure 
\begin{equation}
\label{eq:y}
\zeta(x)
=\frac{k_B\sigma_T}{m_e c^2}n_e(x)T_e(x) \:.
\end{equation}
Typical parameters are $z_{max}=2$, $M_{min}=10^{13}h^{-1}M_{\odot}$ and 
$M_{max}=10^{16}h^{-1}M_{\odot}$. The concentration parameter is calculated
as above, using our merger-tree model of cluster evolution. The temperature and
density profiles are given by eq. (\ref{eq:rhogas})-(\ref{eq:tgas}).

\begin{figure}
\centering
\epsfig{file=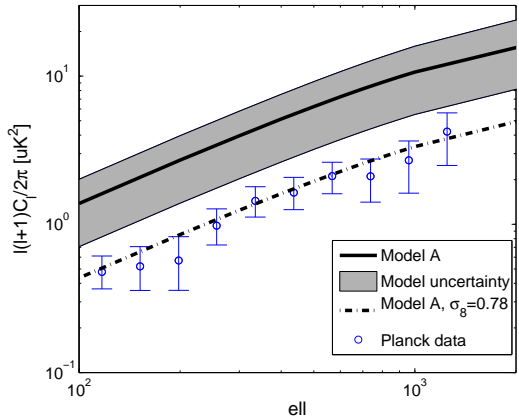, height=6cm}
\caption{SZ power spectrum for the fiducial model (black line) and the model
uncertainty region (grey stripe) that corresponds to the blue lines on Fig.
\ref{fig:fgas_main}.
Also shown are measurements from \emph{Planck} (blue circles), and the
fiducial model with $\sigma_8=0.78$, which corresponds to the lower $2\sigma$
limit of the constraint from \emph{Planck} (dot-dashed black line).}
\label{fig:fgas_sz}
\end{figure}

The resulting thermal SZ power spectrum is shown in Fig. \ref{fig:fgas_sz}.
Given the cosmological parameters deduced from primary CMB observations by
\emph{Planck} (adopted in this work) 
our fiducial model is in tension with \emph{Planck} measurements of the SZ
power 
spectrum. This result reflects the $\sim 2\sigma$ discrepancy between the values
of 
$\sigma_8$ and $\Omega_m$ deduced from primary CMB vs. cluster number counts
\citep{2014A&A...571A..20P}. 
We demonstrate the important implication of this discrepancy by the dashed black
line in 
Fig. \ref{fig:fgas_sz}, calculated with our fiducial $f_{gas}$ model and
$\sigma_8=0.78$, 
which corresponds to the $2\sigma$ lower limit of the \emph{Planck} primary CMB 
value. It is apparent, then, that the uncertainty in the value of $\sigma_8$ can
largely explain the discrepancy with the deduced SZ power spectrum, even if
other additional uncertainties in cluster parameters (such as the mass function
and the gas equation of state) are ignored.

\section{Discussion}
\label{sec:discussion}

We developed a simple and efficient model that accurately describes the mass
dependence of the hot gas mass fraction in clusters. Our model links two
important physical phenomena: the morphological transformation (and mass loss)
of cluster galaxies under the influence of the dense cluster environment, and
the
evolution of the hot gas. Our results show that the possible relation between
these processes can be understood in terms of a few parameters with 
intuitive physical interpretation: The amount of galaxies,
the gas fraction of diffuse matter, and the efficiency of gas removal
from galaxies which we modeled as a power-law in 
halo mass. At present, none of these parameters is known with high precision;
detailed hydrodynamical simulations are needed in order to 
determine the properties of high-redshift galaxies and 
to better understand the IC processes that affect the evolution of their IS gas.

However, our model offers an alternative way to understand IC gas evolution
through
comparison with the observed $f_{gas}$ and metallicity. While it is
computationally 
challenging to develop and run hydrodynamical simulations of cosmological
structure formation that also resolve structure and (relatively) small-scale
galactic processes, such as ram pressure stripping, our
approach provides a convenient framework for 
studying the important gas ejection processes. 
A general model can be derived by fitting $f_{gas}(M)$ measurements, as 
demonstrated above. This model predicts specific dependence of the efficiency
of 
gas removal from galaxies, which can be tested against small-scale numerical 
simulations of ram pressure and tidal 
stripping, processes whose quantitative assessment does not require the full
framework 
of a cosmological simulation. 
Such simulations can be run with different galactic masses and ambient
IC gas densities to study how these parameters contribute to the scatter in
$f_{m,0}$ and $\alpha$ that control gas removal efficiency in our model.
Additional constraints can be provided by measurements of IC gas metallicity,
as demonstrated in the previous section.

An important and timely application of our simple numerical approach 
is the prediction of the SZ power spectrum. 
We reproduce the $\sim 2\sigma$ discrepancy between the models based on 
cosmological parameters deduced from primary CMB observations, and the 
observationally deduced (by \emph{Planck})
thermal SZ power spectrum. While 
the range of model parameters adopted here do not seem to resolve this
discrepancy,
the insight gained from our treatment can be useful 
in future studies of the SZ 
effect, which will allow better assessment of the 
uncertainties resulting from IC gas physics.

It is a well-known fact that preheating models which assume a uniform
entropy floor provide a better description of the observed
luminosity-temperature relation than the simple polytropic model. Nevertheless,
the nature of the preheating sources is still debated. Interestingly, the 
degree of preheating and its impact on 
the equation of state might be related to the IC gas
enrichment processes we discuss here. In particular, if preheating occurs due to
galactic winds, which also carry metals, we would expect a correlation between
the entropy floor and metal abundance, whereas if the main preheating source is
feedback from AGN no such correlation is expected. In addition, non-uniform
preheating should influence the gas mass fraction by 
expelling 
the gas out of less massive systems. This possibility is explored in
\citet{2005ApJ...634..964O} and \citet{2009ApJ...700..989B}. In future work we
plan to extend our model to account for energy, as well as mass ejection from
cluster galaxies, and to use the framework developed in this paper to
distinguish between different 
preheating 
scenarios.

We plan to further extend this work by implementing different mass functions and
thereby
providing much more accurate calculations of the 
cluster statistical measures. 
Another route of investigation is the introduction of cluster-to-cluster scatter
in our 
model parameters. The ultimate goal is constraining our model parameters and
their 
scatter using the observables discussed above - $f_{gas}(M)$, metallicity,
X-ray 
scaling relations and luminosity function, and SZ power spectrum - and
providing 
a handle on the details of the environmental processes that mostly affect
cluster 
galaxies. These results can be used as an input, or compared against,
small-scale
simulations of galaxies in an ambient gas environment, where the modeling of
galactic structure and 
dynamics inside the cluster can be controlled.

\section*{Acknowledgments}                                                      

The authors wish to thank the \textsc{galform} team for making the code publicly
available. We thank the referee for a critical review of the original manuscript
and for constructive comments; we also thank Joe Silk for helpful suggestions. 
Work at TAU was supported by Israel Science Foundation grant 1496/12, and by 
the James B. Ax Family Foundation. Work
at IAP has been supported by the ERC Project No. 267117 (DARK) hosted by
Universit\'{e} Pierre et Marie Curie (UPMC) - Paris 6.

\bibliographystyle{mn2e}
\bibliography{fgas}

\begin{thebibliography}{85}
\expandafter\ifx\csname natexlab\endcsname\relax\def\natexlab#1{#1}\fi

\bibitem[{{Abadi}, {Moore} \& {Bower}(1999){Abadi}, {Moore}, \&
  {Bower}}]{1999MNRAS.308..947A}
{Abadi} M.~G., {Moore} B., {Bower} R.~G., 1999, \mnras, 308, 947

\bibitem[{{Allen} {et~al}\mbox{.}(2008){Allen}, {Rapetti}, {Schmidt},
  {Ebeling}, {Morris}, \& {Fabian}}]{2008MNRAS.383..879A}
{Allen} S.~W., {Rapetti} D.~A., {Schmidt} R.~W., {Ebeling} H., {Morris} R.~G.,
  {Fabian} A.~C., 2008, \mnras, 383, 879

\bibitem[{{Arieli}, {Rephaeli} \& {Norman}(2010){Arieli}, {Rephaeli}, \&
  {Norman}}]{2010ApJ...716..918A}
{Arieli} Y., {Rephaeli} Y., {Norman} M.~L., 2010, \apj, 716, 918

\bibitem[{{Ascasibar} {et~al}\mbox{.}(2003){Ascasibar}, {Yepes}, {M{\"u}ller},
  \& {Gottl{\"o}ber}}]{2003MNRAS.346..731A}
{Ascasibar} Y., {Yepes} G., {M{\"u}ller} V., {Gottl{\"o}ber} S., 2003, \mnras,
  346, 731

\bibitem[{{Babul} {et~al}\mbox{.}(2002){Babul}, {Balogh}, {Lewis}, \&
  {Poole}}]{2002MNRAS.330..329B}
{Babul} A., {Balogh} M.~L., {Lewis} G.~F., {Poole} G.~B., 2002, \mnras, 330,
  329

\bibitem[{{Baldi} {et~al}\mbox{.}(2012){Baldi}, {Ettori}, {Molendi},
  {Balestra}, {Gastaldello}, \& {Tozzi}}]{2012A&A...537A.142B}
{Baldi} A., {Ettori} S., {Molendi} S., {Balestra} I., {Gastaldello} F., {Tozzi}
  P., 2012, \aap, 537, A142

\bibitem[{{Bekki}, {Couch} \& {Shioya}(2002){Bekki}, {Couch}, \&
  {Shioya}}]{2002ApJ...577..651B}
{Bekki} K., {Couch} W.~J., {Shioya} Y., 2002, \apj, 577, 651

\bibitem[{{Bialek}, {Evrard} \& {Mohr}(2001){Bialek}, {Evrard}, \&
  {Mohr}}]{2001ApJ...555..597B}
{Bialek} J.~J., {Evrard} A.~E., {Mohr} J.~J., 2001, \apj, 555, 597

\bibitem[{{B{\^i}rzan} {et~al}\mbox{.}(2004){B{\^i}rzan}, {Rafferty},
  {McNamara}, {Wise}, \& {Nulsen}}]{2004ApJ...607..800B}
{B{\^i}rzan} L., {Rafferty} D.~A., {McNamara} B.~R., {Wise} M.~W., {Nulsen}
  P.~E.~J., 2004, \apj, 607, 800

\bibitem[{{Bode}, {Ostriker} \& {Vikhlinin}(2009){Bode}, {Ostriker}, \&
  {Vikhlinin}}]{2009ApJ...700..989B}
{Bode} P., {Ostriker} J.~P., {Vikhlinin} A., 2009, \apj, 700, 989

\bibitem[{{B{\"o}hringer}, {Chon} \& {Collins}(2014){B{\"o}hringer}, {Chon}, \&
  {Collins}}]{2014A&A...570A..31B}
{B{\"o}hringer} H., {Chon} G., {Collins} C.~A., 2014, \aap, 570, A31

\bibitem[{{Borgani} {et~al}\mbox{.}(2006){Borgani}, {Dolag}, {Murante},
  {Cheng}, {Springel}, {Diaferio}, {Moscardini}, {Tormen}, {Tornatore}, \&
  {Tozzi}}]{2006MNRAS.367.1641B}
{Borgani} S. {et~al.}, 2006, \mnras, 367, 1641

\bibitem[{{Br{\"u}ggen} \& {De Lucia}(2008)}]{2008MNRAS.383.1336B}
{Br{\"u}ggen} M., {De Lucia} G., 2008, \mnras, 383, 1336

\bibitem[{{Bullock} {et~al}\mbox{.}(2001){Bullock}, {Kolatt}, {Sigad},
  {Somerville}, {Kravtsov}, {Klypin}, {Primack}, \&
  {Dekel}}]{2001MNRAS.321..559B}
{Bullock} J.~S., {Kolatt} T.~S., {Sigad} Y., {Somerville} R.~S., {Kravtsov}
  A.~V., {Klypin} A.~A., {Primack} J.~R., {Dekel} A., 2001, \mnras, 321, 559

\bibitem[{{Butcher} \& {Oemler}(1978)}]{1978ApJ...219...18B}
{Butcher} H., {Oemler}, Jr. A., 1978, \apj, 219, 18

\bibitem[{{Dai} {et~al}\mbox{.}(2010){Dai}, {Bregman}, {Kochanek}, \&
  {Rasia}}]{2010ApJ...719..119D}
{Dai} X., {Bregman} J.~N., {Kochanek} C.~S., {Rasia} E., 2010, \apj, 719, 119

\bibitem[{{Dalal}, {Lithwick} \& {Kuhlen}(2010){Dalal}, {Lithwick}, \&
  {Kuhlen}}]{2010arXiv1010.2539D}
{Dalal} N., {Lithwick} Y., {Kuhlen} M., 2010, preprint (arXiv:1010.2539)

\bibitem[{{De Grandi} \& {Molendi}(2001)}]{2001ApJ...551..153D}
{De Grandi} S., {Molendi} S., 2001, \apj, 551, 153

\bibitem[{{Dunn}, {Fabian} \& {Taylor}(2005){Dunn}, {Fabian}, \&
  {Taylor}}]{2005MNRAS.364.1343D}
{Dunn} R.~J.~H., {Fabian} A.~C., {Taylor} G.~B., 2005, \mnras, 364, 1343

\bibitem[{{Dvorkin} \& {Rephaeli}(2011)}]{2011MNRAS.412..665D}
{Dvorkin} I., {Rephaeli} Y., 2011, \mnras, 412, 665

\bibitem[{{Dvorkin}, {Rephaeli} \& {Shimon}(2012){Dvorkin}, {Rephaeli}, \&
  {Shimon}}]{2012MNRAS.421.2648D}
{Dvorkin} I., {Rephaeli} Y., {Shimon} M., 2012, \mnras, 421, 2648

\bibitem[{{Ebeling}, {Stephenson} \& {Edge}(2014){Ebeling}, {Stephenson}, \&
  {Edge}}]{2014ApJ...781L..40E}
{Ebeling} H., {Stephenson} L.~N., {Edge} A.~C., 2014, \apjl, 781, L40

\bibitem[{{Ettori}(2003)}]{2003MNRAS.344L..13E}
{Ettori} S., 2003, \mnras, 344, L13

\bibitem[{{Ettori} {et~al}\mbox{.}(2006){Ettori}, {Dolag}, {Borgani}, \&
  {Murante}}]{2006MNRAS.365.1021E}
{Ettori} S., {Dolag} K., {Borgani} S., {Murante} G., 2006, \mnras, 365, 1021

\bibitem[{{Fabian}(2012)}]{2012ARA&A..50..455F}
{Fabian} A.~C., 2012, \araa, 50, 455

\bibitem[{{Finoguenov}, {David} \& {Ponman}(2000){Finoguenov}, {David}, \&
  {Ponman}}]{2000ApJ...544..188F}
{Finoguenov} A., {David} L.~P., {Ponman} T.~J., 2000, \apj, 544, 188

\bibitem[{{Giodini} {et~al}\mbox{.}(2009){Giodini}, {Pierini}, {Finoguenov},
  {Pratt}, {Boehringer}, {Leauthaud}, {Guzzo}, {Aussel}, {Bolzonella}, {Capak},
  {Elvis}, {Hasinger}, {Ilbert}, {Kartaltepe}, {Koekemoer}, {Lilly}, {Massey},
  {McCracken}, {Rhodes}, {Salvato}, {Sanders}, {Scoville}, {Sasaki}, {Smolcic},
  {Taniguchi}, {Thompson}, \& {COSMOS Collaboration}}]{2009ApJ...703..982G}
{Giodini} S. {et~al.}, 2009, \apj, 703, 982

\bibitem[{{Gnedin}(2003)}]{2003ApJ...582..141G}
{Gnedin} O.~Y., 2003, \apj, 582, 141

\bibitem[{{Gonzalez} {et~al}\mbox{.}(2013){Gonzalez}, {Sivanandam},
  {Zabludoff}, \& {Zaritsky}}]{2013ApJ...778...14G}
{Gonzalez} A.~H., {Sivanandam} S., {Zabludoff} A.~I., {Zaritsky} D., 2013,
  \apj, 778, 14

\bibitem[{{Gonzalez}, {Zaritsky} \& {Zabludoff}(2007){Gonzalez}, {Zaritsky}, \&
  {Zabludoff}}]{2007ApJ...666..147G}
{Gonzalez} A.~H., {Zaritsky} D., {Zabludoff} A.~I., 2007, \apj, 666, 147

\bibitem[{{Gunn} \& {Gott}(1972)}]{1972ApJ...176....1G}
{Gunn} J.~E., {Gott}, III J.~R., 1972, \apj, 176, 1

\bibitem[{{Hester}(2006)}]{2006ApJ...647..910H}
{Hester} J.~A., 2006, \apj, 647, 910

\bibitem[{{Hlavacek-Larrondo} {et~al}\mbox{.}(2012){Hlavacek-Larrondo},
  {Fabian}, {Edge}, {Ebeling}, {Sanders}, {Hogan}, \&
  {Taylor}}]{2012MNRAS.421.1360H}
{Hlavacek-Larrondo} J., {Fabian} A.~C., {Edge} A.~C., {Ebeling} H., {Sanders}
  J.~S., {Hogan} M.~T., {Taylor} G.~B., 2012, \mnras, 421, 1360

\bibitem[{{Hlavacek-Larrondo} {et~al}\mbox{.}(2014){Hlavacek-Larrondo},
  {McDonald}, {Benson}, {Forman}, {Allen}, {Bleem}, {Ashby}, {Bocquet},
  {Brodwin}, {Dietrich}, {Jones}, {Liu}, {Saliwanchik}, {Saro}, {Schrabback},
  {Song}, {Stalder}, {Vikhlinin}, \& {Zenteno}}]{2014arXiv1410.0025H}
{Hlavacek-Larrondo} J. {et~al.}, 2014, preprint (arXiv:1410.0025)

\bibitem[{{Hughes} \& {Cortese}(2009)}]{2009MNRAS.396L..41H}
{Hughes} T.~M., {Cortese} L., 2009, \mnras, 396, L41

\bibitem[{{Just} {et~al}\mbox{.}(2010){Just}, {Zaritsky}, {Sand}, {Desai}, \&
  {Rudnick}}]{2010ApJ...711..192J}
{Just} D.~W., {Zaritsky} D., {Sand} D.~J., {Desai} V., {Rudnick} G., 2010,
  \apj, 711, 192

\bibitem[{{Kapferer} {et~al}\mbox{.}(2007){Kapferer}, {Kronberger},
  {Weratschnig}, {Schindler}, {Domainko}, {van Kampen}, {Kimeswenger}, {Mair},
  \& {Ruffert}}]{2007A&A...466..813K}
{Kapferer} W. {et~al.}, 2007, \aap, 466, 813

\bibitem[{{Komatsu} \& {Seljak}(2002)}]{2002MNRAS.336.1256K}
{Komatsu} E., {Seljak} U., 2002, \mnras, 336, 1256

\bibitem[{{Koopmann}, {Haynes} \& {Catinella}(2006){Koopmann}, {Haynes}, \&
  {Catinella}}]{2006AJ....131..716K}
{Koopmann} R.~A., {Haynes} M.~P., {Catinella} B., 2006, \aj, 131, 716

\bibitem[{{Lacey} \& {Cole}(1993)}]{1993MNRAS.262..627L}
{Lacey} C., {Cole} S., 1993, \mnras, 262, 627

\bibitem[{{Limousin} {et~al}\mbox{.}(2009){Limousin}, {Sommer-Larsen},
  {Natarajan}, \& {Milvang-Jensen}}]{2009ApJ...696.1771L}
{Limousin} M., {Sommer-Larsen} J., {Natarajan} P., {Milvang-Jensen} B., 2009,
  \apj, 696, 1771

\bibitem[{{Lin}, {Mohr} \& {Stanford}(2003){Lin}, {Mohr}, \&
  {Stanford}}]{2003ApJ...591..749L}
{Lin} Y.-T., {Mohr} J.~J., {Stanford} S.~A., 2003, \apj, 591, 749

\bibitem[{{Lueker} {et~al}\mbox{.}(2010){Lueker}, {Reichardt}, {Schaffer},
  {Zahn}, {Ade}, {Aird}, {Benson}, {Bleem}, {Carlstrom}, {Chang}, {Cho},
  {Crawford}, {Crites}, {de Haan}, {Dobbs}, {George}, {Hall}, {Halverson},
  {Holder}, {Holzapfel}, {Hrubes}, {Joy}, {Keisler}, {Knox}, {Lee}, {Leitch},
  {McMahon}, {Mehl}, {Meyer}, {Mohr}, {Montroy}, {Padin}, {Plagge}, {Pryke},
  {Ruhl}, {Shaw}, {Shirokoff}, {Spieler}, {Stalder}, {Staniszewski}, {Stark},
  {Vanderlinde}, {Vieira}, \& {Williamson}}]{2010ApJ...719.1045L}
{Lueker} M. {et~al.}, 2010, \apj, 719, 1045

\bibitem[{{Mantz} {et~al}\mbox{.}(2010{\natexlab{a}}){Mantz}, {Allen},
  {Ebeling}, {Rapetti}, \& {Drlica-Wagner}}]{2010MNRAS.406.1773M}
{Mantz} A., {Allen} S.~W., {Ebeling} H., {Rapetti} D., {Drlica-Wagner} A.,
  2010{\natexlab{a}}, \mnras, 406, 1773

\bibitem[{{Mantz} {et~al}\mbox{.}(2010{\natexlab{b}}){Mantz}, {Allen},
  {Rapetti}, \& {Ebeling}}]{2010MNRAS.406.1759M}
{Mantz} A., {Allen} S.~W., {Rapetti} D., {Ebeling} H., 2010{\natexlab{b}},
  \mnras, 406, 1759

\bibitem[{{McCarthy}, {Bower} \& {Balogh}(2007){McCarthy}, {Bower}, \&
  {Balogh}}]{2007MNRAS.377.1457M}
{McCarthy} I.~G., {Bower} R.~G., {Balogh} M.~L., 2007, \mnras, 377, 1457

\bibitem[{{McDonald} {et~al}\mbox{.}(2014){McDonald}, {Benson}, {Vikhlinin},
  {Aird}, {Allen}, {Bautz}, {Bayliss}, {Bleem}, {Bocquet}, {Brodwin},
  {Carlstrom}, {Chang}, {Cho}, {Clocchiatti}, {Crawford}, {Crites}, {de Haan},
  {Dobbs}, {Foley}, {Forman}, {George}, {Gladders}, {Gonzalez}, {Halverson},
  {Hlavacek-Larrondo}, {Holder}, {Holzapfel}, {Hrubes}, {Jones}, {Keisler},
  {Knox}, {Lee}, {Leitch}, {Liu}, {Lueker}, {Luong-Van}, {Mantz}, {Marrone},
  {McMahon}, {Meyer}, {Miller}, {Mocanu}, {Mohr}, {Murray}, {Padin}, {Pryke},
  {Reichardt}, {Rest}, {Ruhl}, {Saliwanchik}, {Saro}, {Sayre}, {Schaffer},
  {Shirokoff}, {Spieler}, {Stalder}, {Stanford}, {Staniszewski}, {Stark},
  {Story}, {Stubbs}, {Vanderlinde}, {Vieira}, {Williamson}, {Zahn}, \&
  {Zenteno}}]{2014ApJ...794...67M}
{McDonald} M. {et~al.}, 2014, \apj, 794, 67

\bibitem[{{McNamara} \& {Nulsen}(2007)}]{2007ARA&A..45..117M}
{McNamara} B.~R., {Nulsen} P.~E.~J., 2007, \araa, 45, 117

\bibitem[{{Merritt}(1983)}]{1983ApJ...264...24M}
{Merritt} D., 1983, \apj, 264, 24

\bibitem[{{Mohr}, {Mathiesen} \& {Evrard}(1999){Mohr}, {Mathiesen}, \&
  {Evrard}}]{1999ApJ...517..627M}
{Mohr} J.~J., {Mathiesen} B., {Evrard} A.~E., 1999, \apj, 517, 627

\bibitem[{{Moore} {et~al}\mbox{.}(1999){Moore}, {Lake}, {Quinn}, \&
  {Stadel}}]{1999MNRAS.304..465M}
{Moore} B., {Lake} G., {Quinn} T., {Stadel} J., 1999, \mnras, 304, 465

\bibitem[{{Ostriker}, {Bode} \& {Babul}(2005){Ostriker}, {Bode}, \&
  {Babul}}]{2005ApJ...634..964O}
{Ostriker} J.~P., {Bode} P., {Babul} A., 2005, \apj, 634, 964

\bibitem[{{Parkinson}, {Cole} \& {Helly}(2008){Parkinson}, {Cole}, \&
  {Helly}}]{2008MNRAS.383..557P}
{Parkinson} H., {Cole} S., {Helly} J., 2008, \mnras, 383, 557

\bibitem[{{Peng} \& {Maiolino}(2014)}]{2014MNRAS.438..262P}
{Peng} Y.-j., {Maiolino} R., 2014, \mnras, 438, 262

\bibitem[{{Pettini} {et~al}\mbox{.}(2001){Pettini}, {Shapley}, {Steidel},
  {Cuby}, {Dickinson}, {Moorwood}, {Adelberger}, \&
  {Giavalisco}}]{2001ApJ...554..981P}
{Pettini} M., {Shapley} A.~E., {Steidel} C.~C., {Cuby} J.-G., {Dickinson} M.,
  {Moorwood} A.~F.~M., {Adelberger} K.~L., {Giavalisco} M., 2001, \apj, 554,
  981

\bibitem[{{Planck Collaboration} {et~al}\mbox{.}(2014{\natexlab{a}}){Planck
  Collaboration}, {Ade}, {Aghanim}, {Armitage-Caplan}, {Arnaud}, {Ashdown},
  {Atrio-Barandela}, {Aumont}, {Baccigalupi}, {Banday}, \&
  et~al.}]{2014A&A...571A..16P}
{Planck Collaboration} {et~al.}, 2014{\natexlab{a}}, \aap, 571, A16

\bibitem[{{Planck Collaboration} {et~al}\mbox{.}(2014{\natexlab{b}}){Planck
  Collaboration}, {Ade}, {Aghanim}, {Armitage-Caplan}, {Arnaud}, {Ashdown},
  {Atrio-Barandela}, {Aumont}, {Baccigalupi}, {Banday}, \&
  et~al.}]{2014A&A...571A..20P}
{Planck Collaboration} {et~al.}, 2014{\natexlab{b}}, \aap, 571, A20

\bibitem[{{Planelles} {et~al}\mbox{.}(2013){Planelles}, {Borgani}, {Dolag},
  {Ettori}, {Fabjan}, {Murante}, \& {Tornatore}}]{2013MNRAS.tmp..918P}
{Planelles} S., {Borgani} S., {Dolag} K., {Ettori} S., {Fabjan} D., {Murante}
  G., {Tornatore} L., 2013, \mnras

\bibitem[{{Pratt} {et~al}\mbox{.}(2010){Pratt}, {Arnaud}, {Piffaretti},
  {B{\"o}hringer}, {Ponman}, {Croston}, {Voit}, {Borgani}, \&
  {Bower}}]{2010A&A...511A..85P}
{Pratt} G.~W. {et~al.}, 2010, \aap, 511, A85

\bibitem[{{Quilis}, {Moore} \& {Bower}(2000){Quilis}, {Moore}, \&
  {Bower}}]{2000Sci...288.1617Q}
{Quilis} V., {Moore} B., {Bower} R., 2000, Science, 288, 1617

\bibitem[{{Raichoor} \& {Andreon}(2012)}]{2012A&A...543A..19R}
{Raichoor} A., {Andreon} S., 2012, \aap, 543, A19

\bibitem[{{Rasmussen} \& {Ponman}(2009)}]{2009MNRAS.399..239R}
{Rasmussen} J., {Ponman} T.~J., 2009, \mnras, 399, 239

\bibitem[{{Sadeh} \& {Rephaeli}(2008)}]{2008MNRAS.388.1759S}
{Sadeh} S., {Rephaeli} Y., 2008, \mnras, 388, 1759

\bibitem[{{Sasaki}, {Matsushita} \& {Sato}(2014){Sasaki}, {Matsushita}, \&
  {Sato}}]{2014ApJ...781...36S}
{Sasaki} T., {Matsushita} K., {Sato} K., 2014, \apj, 781, 36

\bibitem[{{Scannapieco} \& {Oh}(2004)}]{2004ApJ...608...62S}
{Scannapieco} E., {Oh} S.~P., 2004, \apj, 608, 62

\bibitem[{{Sheth} \& {Tormen}(1999)}]{1999MNRAS.308..119S}
{Sheth} R.~K., {Tormen} G., 1999, \mnras, 308, 119

\bibitem[{{Shimon} {et~al}\mbox{.}(2012){Shimon}, {Rephaeli}, {Itzhaki},
  {Dvorkin}, \& {Keating}}]{2012MNRAS.427..828S}
{Shimon} M., {Rephaeli} Y., {Itzhaki} N., {Dvorkin} I., {Keating} B.~G., 2012,
  \mnras, 427, 828

\bibitem[{{Solanes} {et~al}\mbox{.}(2001){Solanes}, {Manrique},
  {Garc{\'{\i}}a-G{\'o}mez}, {Gonz{\'a}lez-Casado}, {Giovanelli}, \&
  {Haynes}}]{2001ApJ...548...97S}
{Solanes} J.~M., {Manrique} A., {Garc{\'{\i}}a-G{\'o}mez} C.,
  {Gonz{\'a}lez-Casado} G., {Giovanelli} R., {Haynes} M.~P., 2001, \apj, 548,
  97

\bibitem[{{Stanek} {et~al}\mbox{.}(2010){Stanek}, {Rasia}, {Evrard}, {Pearce},
  \& {Gazzola}}]{2010ApJ...715.1508S}
{Stanek} R., {Rasia} E., {Evrard} A.~E., {Pearce} F., {Gazzola} L., 2010, \apj,
  715, 1508

\bibitem[{{Sun}, {Donahue} \& {Voit}(2007){Sun}, {Donahue}, \&
  {Voit}}]{2007ApJ...671..190S}
{Sun} M., {Donahue} M., {Voit} G.~M., 2007, \apj, 671, 190

\bibitem[{{Sun} {et~al}\mbox{.}(2009){Sun}, {Voit}, {Donahue}, {Jones},
  {Forman}, \& {Vikhlinin}}]{2009ApJ...693.1142S}
{Sun} M., {Voit} G.~M., {Donahue} M., {Jones} C., {Forman} W., {Vikhlinin} A.,
  2009, \apj, 693, 1142

\bibitem[{{Tecce} {et~al}\mbox{.}(2010){Tecce}, {Cora}, {Tissera}, {Abadi}, \&
  {Lagos}}]{2010MNRAS.408.2008T}
{Tecce} T.~E., {Cora} S.~A., {Tissera} P.~B., {Abadi} M.~G., {Lagos} C.~D.~P.,
  2010, \mnras, 408, 2008

\bibitem[{{Vikhlinin} {et~al}\mbox{.}(2006){Vikhlinin}, {Kravtsov}, {Forman},
  {Jones}, {Markevitch}, {Murray}, \& {Van Speybroeck}}]{2006ApJ...640..691V}
{Vikhlinin} A., {Kravtsov} A., {Forman} W., {Jones} C., {Markevitch} M.,
  {Murray} S.~S., {Van Speybroeck} L., 2006, \apj, 640, 691

\bibitem[{{Vikhlinin} {et~al}\mbox{.}(2009){Vikhlinin}, {Kravtsov}, {Burenin},
  {Ebeling}, {Forman}, {Hornstrup}, {Jones}, {Murray}, {Nagai}, {Quintana}, \&
  {Voevodkin}}]{2009ApJ...692.1060V}
{Vikhlinin} A. {et~al.}, 2009, \apj, 692, 1060

\bibitem[{{Vikhlinin} {et~al}\mbox{.}(2005){Vikhlinin}, {Markevitch}, {Murray},
  {Jones}, {Forman}, \& {Van Speybroeck}}]{2005ApJ...628..655V}
{Vikhlinin} A., {Markevitch} M., {Murray} S.~S., {Jones} C., {Forman} W., {Van
  Speybroeck} L., 2005, \apj, 628, 655

\bibitem[{{Voit} {et~al}\mbox{.}(2003){Voit}, {Balogh}, {Bower}, {Lacey}, \&
  {Bryan}}]{2003ApJ...593..272V}
{Voit} G.~M., {Balogh} M.~L., {Bower} R.~G., {Lacey} C.~G., {Bryan} G.~L.,
  2003, \apj, 593, 272

\bibitem[{{Vollmer} {et~al}\mbox{.}(2001){Vollmer}, {Cayatte}, {Balkowski}, \&
  {Duschl}}]{2001ApJ...561..708V}
{Vollmer} B., {Cayatte} V., {Balkowski} C., {Duschl} W.~J., 2001, \apj, 561,
  708

\bibitem[{{Wechsler} {et~al}\mbox{.}(2002){Wechsler}, {Bullock}, {Primack},
  {Kravtsov}, \& {Dekel}}]{2002ApJ...568...52W}
{Wechsler} R.~H., {Bullock} J.~S., {Primack} J.~R., {Kravtsov} A.~V., {Dekel}
  A., 2002, \apj, 568, 52

\bibitem[{{Werner} {et~al}\mbox{.}(2013){Werner}, {Urban}, {Simionescu}, \&
  {Allen}}]{2013Natur.502..656W}
{Werner} N., {Urban} O., {Simionescu} A., {Allen} S.~W., 2013, \nat, 502, 656

\bibitem[{{White} {et~al}\mbox{.}(1993){White}, {Navarro}, {Evrard}, \&
  {Frenk}}]{1993Natur.366..429W}
{White} S.~D.~M., {Navarro} J.~F., {Evrard} A.~E., {Frenk} C.~S., 1993, \nat,
  366, 429

\bibitem[{{Yoo} {et~al}\mbox{.}(2007){Yoo}, {Miralda-Escud{\'e}}, {Weinberg},
  {Zheng}, \& {Morgan}}]{2007ApJ...667..813Y}
{Yoo} J., {Miralda-Escud{\'e}} J., {Weinberg} D.~H., {Zheng} Z., {Morgan}
  C.~W., 2007, \apj, 667, 813

\bibitem[{{Young} {et~al}\mbox{.}(2011){Young}, {Thomas}, {Short}, \&
  {Pearce}}]{2011MNRAS.413..691Y}
{Young} O.~E., {Thomas} P.~A., {Short} C.~J., {Pearce} F., 2011, \mnras, 413,
  691

\bibitem[{{Younger} \& {Bryan}(2007)}]{2007ApJ...666..647Y}
{Younger} J.~D., {Bryan} G.~L., 2007, \apj, 666, 647

\bibitem[{{Zhang} {et~al}\mbox{.}(2008){Zhang}, {Finoguenov}, {B{\"o}hringer},
  {Kneib}, {Smith}, {Kneissl}, {Okabe}, \& {Dahle}}]{2008A&A...482..451Z}
{Zhang} Y.-Y., {Finoguenov} A., {B{\"o}hringer} H., {Kneib} J.-P., {Smith}
  G.~P., {Kneissl} R., {Okabe} N., {Dahle} H., 2008, \aap, 482, 451

\bibitem[{{Zhang} {et~al}\mbox{.}(2011){Zhang}, {Lagan{\'a}}, {Pierini},
  {Puchwein}, {Schneider}, \& {Reiprich}}]{2011A&A...535A..78Z}
{Zhang} Y.-Y., {Lagan{\'a}} T.~F., {Pierini} D., {Puchwein} E., {Schneider} P.,
  {Reiprich} T.~H., 2011, \aap, 535, A78

\end{thebibliography}

\appendix
\section{Preheating model}

Our simple phenomenological preheating model is based on the analytic model of
\citet{2007ApJ...666..647Y}.
As is customary in the literature, we define entropy as 
\begin{equation}
 K=\frac{P}{\rho_g^{\gamma}}\:,
\label{eq:Kdef}
\end{equation}
where $P$ and $\rho_g$ are the pressure and density of the gas and $\gamma$
is the adiabatic index.

For each cluster we start with the polytropic model in hydrostatic equilibrium
described in Section \ref{sec:xray}, calculate the entropy and modify it 
as 
\begin{equation}
 \hat{K}(r)=K(r)+K_0 , 
\end{equation}
thus adding an entropy floor. We then solve the equation of hydrostatic
equilibrium:
\begin{equation}
 \frac{d\hat{P}}{dr}=-\frac{GM_{tot}(r)}{r^2}\hat{\rho}_g(r)\:,
\label{eq:Khydro}
\end{equation}
where modified quantities are denoted by hats, and the density is given by eq.
(\ref{eq:Kdef}). The temperature is then given by:
\begin{equation}
 k_BT=K^{3/5}P^{2/5}\:.
\end{equation}
In solving equation (\ref{eq:Khydro}) we choose the pressure boundary condition
at the virial radius $P(R)$, such that the total gas mass is conserved (no gas
outflow due to increased entropy). Relaxing this assumption will slightly modify
the gas mass fraction, and in this sense the model presented here is not
fully
self-consistent. However, this assumption is expected to be 
reasonably adequate 
if preheating was uniform and took place long before groups and clusters 
formed.

\end{document}